\documentclass[aps,
prl,
twocolumn,
showpacs,
superscriptaddress,
groupedaddress,
floatfix, 10pt]{revtex4-1}  

\usepackage{graphicx}
\usepackage{bm}
\usepackage{slashed}
\pdfoutput=1
\usepackage{commath}
\usepackage{hyperref}

\DeclareMathOperator{\tr}{tr}

\begin{document}
\title{Enriched axial anomaly in Weyl materials}

\author{Zachary M. Raines}
\author{Victor M. Galitski}
\affiliation{Joint Quantum Institute and Department of Physics, University of Maryland, College Park, Maryland 20742-4111, U.S.A.} 
\begin{abstract}
    While quantum anomalies are often associated with the breaking of a classical symmetry in the quantum theory, their anomalous contributions to observables remain distinct and well-defined even when the symmetry is broken from the outset.
This paper explores such anomalous contributions to the current, originating from the axial anomaly in a Weyl semimetal, and in the presence of a generic Weyl node-mixing term.
We find that apart from the familiar anomalous divergence of the axial current proportional to a product of electric and magnetic fields, there is another anomalous term proportional to a product of the electric field and the orientation of a spin-dependent node-mixing vector.
We obtain this result both by a quantum field-theoretic analysis of an effective Weyl action and solving an explicit lattice model.
The extended spin-mixing mass terms, and the enriched axial anomaly they entail,  could arise as mean-field or proximity-induced order parameters in spin-density-wave phases in Weyl semimetals or be generated dynamically within a Floquet theory.
\end{abstract}
\maketitle

Quantum anomalies represent a surprising deviation from classical intuition, where a global symmetry of the classical Lagrangian does not necessarily lead to a conservation law of the corresponding charge. This fact entails profound and fundamental consequences in particle physics such as non-conservation of baryon charge and the appearance of instantons and $\theta$-vacuum in quantum chromodynamics.

Anomalies, including the chiral or axial anomaly~\cite{Adler1969,Bell1969,Nielsen1983,Fujikawa1984}, also appear in the field-theoretic descriptions of condensed matter models and give rise to ``anomalous'' contributions to observable currents. Most recently, the condensed-matter chiral anomaly has been discussed extensively in the context of Weyl semimetals, and its signatures were experimentally observed in magneto-transport of these materials~\cite{Huang2015}.

Arguably, the condensed matter axial anomalies are less impressive than in particle physics as far as symmetry-breaking is concerned. They oftentimes represent properties and ``quantum symmetry breaking'' of a low-energy effective theory, while the origin of the anomaly derives from a short-distance regularization, where the low-energy theory is not quantitatively applicable and chiral symmetry is poorly defined. Once the full theory is restored, the symmetry of the low-energy model is no longer tied to strong conservation laws. For example, the non-conservation of charge attached to a particular Weyl node in a Weyl semimetal is not particularly surprising once we recall that the Weyl nodes are connected through the bottom of the band in the full lattice band structure~\cite{Nielsen1983,Son2013}. While the breaking of the chiral symmetry in condensed matter systems is not unexpected, the anomalous contributions predicted within the low energy theory remain observable effects, for example the anomalous Hall and chiral magnetic effects in Weyl semimetals.

In fact, these ``chiral-anomalous'' contributions and features generally survive even if there is no chiral symmetry to be broken even in the low-energy model. One concrete way to express an anomaly in such a context is by comparing the divergence of the classical Noether current \textit{vs} the associated Ward identity obtained from the quantum theory. For example, even if Dirac mass terms are included in a description of a Weyl semimetal (which physically implies scattering between the Weyl nodes that breaks chiral symmetry already at the classical level), one can still identify a well-defined anomalous contribution, as was considered for example by Zyuzin and Burkov in Ref.~[\onlinecite{Zyuzin2012}]. In this case, one finds classically that $\partial_\mu j^\mu_5 = -2 i m \bar\psi \gamma^5\psi$,
whereas in the quantum theory this is modified to become $\partial_\mu \langle j^\mu_5\rangle = -2 i m \langle \bar\psi \gamma^5\psi\rangle + {\cal A}(x)$.
The presence of the anomaly function ${\cal A}(x)=(e^2/2\pi^2) \mathbf{E} \cdot \mathbf{B}$ is a hallmark of a quantum anomaly, which persists even though the symmetry corresponding to $j^\mu_5$ is already broken at the classical level by the presence of a Dirac mass.

Motivated by these considerations, we explore the anomalous divergence of the Noether current in the presence of a generic node-mixing term in a Weyl semimetal. Our main result, derived and discussed below, is that the spin-dependent node-mixing terms do not affect the ``conventional'' chiral anomaly, but give rise to another anomalous term \textemdash\ $\mathcal{A}_\text{spin} (x) = -\frac{e}{\pi} \mathbf{E} \cdot \mathrm{Re}\, \left( \mathbf{g} m^* \right)$, where the vector $\mathbf{g}$ defined in Eq.~\eqref{eq:g} below, determines spin-mixing between the nodes and $m$ is the complex Dirac mass.
As is the case with the conventional anomaly, the presence of such an anomaly term can lead to anomalous transport, which in this case is reminiscent of the chiral magnetic effect.


We will begin by examining a low-energy theory of Dirac fermions within the functional integral technique.
Our aim is to understand what electromagnetic response can arise from the addition of new terms to the Lagrangian.
To do so we employ the chiral rotation technique as first illustrated by Fujikawa~\cite{Fujikawa1984}. Our starting point is the Dirac Lagrangian in imaginary time
\begin{equation}
    \mathcal{L} =
    \bar{\psi}
    \left[
        i \slashed{D}  - \slashed{b} \gamma^5 - |m|e^{i\alpha\gamma^5}
        - \Delta_{\mu\nu} \sigma^{\mu \nu}
        \right]\psi
        \equiv
    i \bar{\psi}\mathcal{D}\psi,
    \label{eq:dirac}
\end{equation}
where $\bar{\psi} = \psi^\dagger \gamma^0$ denotes Dirac conjugation and we are using Euclidean gamma matrices $\{\gamma^\mu, \gamma^\nu\} = 2 g^{\mu\nu} = - 2 \delta^{\mu\nu}$, with $\mu,\nu = 1,2,3,4$.
We use the Feynman slash notation $\slashed{D} = D_\mu \gamma^\mu$ where $D_\mu = \partial_\mu + i e A_\mu$ is the gauge covariant derivative.
We also define the fifth gamma matrix $\gamma^5 = \gamma^1 \gamma^2 \gamma^3 \gamma^4$ and the sigma matrices $\sigma^{\mu\nu} = (i/2)[\gamma^\mu, \gamma^\nu]$.
With these definitions the matrices $\gamma^\mu$ are anti-Hermitian, while $\gamma^5$ and $\sigma^{\mu\nu}$ are Hermitian.
We defer a discussion of the specific origin of $m$ and $\Delta$ until later in the paper.

We now perform a chiral gauge transformation to remove the axial vector $b^\mu$ from the fermionic action.
Subsequently, we make the following claims in the absence of an external field.
Firstly, the fermionic sector exhibits no axial-vector dependent currents.
Secondly, the Jacobian introduced via the anomaly in the path integral measure produces a current in response to the $\Delta_{\mu\nu}$ term.

As pointed out by Fujikawa, when considering the chiral anomaly it is important to specify which basis one is using to define the functional integral~\cite{Fujikawa1984}.
In his original work, Fujikawa used the basis states of the Euclidean operator $\slashed{D}$ which is Hermitian.
Here, the operator $\mathcal{D}$ is not Hermitian, and so we follow the approach of Refs.~[\onlinecite{Kikuchi1992,Zyuzin2012}].
We define the eigenfunctions and eigenvalues of the operators $\mathcal{D}^\dagger\mathcal{D}$ and $\mathcal{D}\mathcal{D}^\dagger$ by
\begin{equation}
\begin{gathered}
    \mathcal{D}^\dagger \mathcal{D}\phi_n(x) = \lambda_n^2 \phi_n(x) \text{ and }
    \tilde{\phi}^\dagger_n(x)\mathcal{D} \mathcal{D}^\dagger = \tilde{\phi}^\dagger_n(x)\lambda_n^2.
\end{gathered}
\end{equation}
These operators are manifestly Hermitian and thus their eigenvectors form a basis.
In defining the eigenvalues as $\lambda_n^2$ we have made use of the fact that the operators $\mathcal{D}^\dagger \mathcal{D}$ and $\mathcal{D} \mathcal{D}^\dagger$ are positive semi-definite.
Note that there is a one-to-one correspondence between their non-zero eigenvalues.

We then define the path-integral by expressing $\psi$ and $\bar \psi$ in terms of the eigenfunctions of $\mathcal{D}^\dagger \mathcal{D}$ and $\mathcal{D} \mathcal{D}^\dagger$ respectively as
\begin{equation}
    \psi(x) = \sum_n \phi_n(x) a_n,\quad
    \bar\psi(x) = \sum_n \bar a_n \tilde{\phi}^\dagger_n(x).
\end{equation}
With our basis states defined we now consider an infinitesimal chiral gauge transformation of the form.
\begin{equation}
\begin{gathered}
    \psi(x) = e^{-i \dif{s}\,b \cdot x \gamma^5} \psi'(x) \text{ and }
    \bar\psi(x) = \bar \psi'(x)e^{-i \dif{s}\,b \cdot x \gamma^5}.
    \label{eq:chiral}
\end{gathered}
\end{equation}
Under such a transformation the Lagrangian becomes
\begin{multline}
    \mathcal{L}' =
    \bar{\psi}'
    \left[
        i \slashed{D}  - (1 - \dif{s})\slashed{b} \gamma^5
        - |m|e^{i(\alpha - 2\dif{s}\,b\cdot x)\gamma^5}\right.\\
       \left. - \Delta_{\mu\nu}e^{-2i\dif{s}\, b \cdot x \gamma^5} \sigma^{\mu \nu}
    \right]\psi'.
    \label{eq:rotl}
\end{multline}
We must also include the contribution from the Jacobian factor in the path integral measure which is introduced by this transformation.
To that end, let us consider the partition function, which under the chiral rotation Eq.~\eqref{eq:chiral} transforms as
\begin{equation}
    Z = \int \mathcal{D}[\bar a, a] e^{-S[\bar a, a]}
    = \int \mathcal{D}[\bar a', a']{(\det J)}^{-1} e^{-S'[\bar a', a']}
\end{equation}
where $S$ is the action corresponding to the Dirac Lagrangian Eq.~\eqref{eq:dirac}, expressed in terms of the basis states $a_n$, and
$S'$ is the action corresponding to the Lagrangian Eq.~\eqref{eq:rotl}.
The new fields of integration $a'_n$ and $\bar a'_n$ are implicitly defined in terms of the old via the relations
\begin{equation}
    a_n = \sum_m U_{nm} a'_m,\quad
    \bar a_n = \sum_m \bar a'_m \tilde{U}_{mn}
\end{equation}
with $U_{nm} = \int dx \phi_n^\dagger (x) e^{-i \dif{s}\, b\cdot x\gamma^5}\phi_m(x)$ and $\widetilde{U}_{nm} = \int dx \tilde{\phi}^\dagger_n(x) e^{-i \dif{s}\, b \cdot x \gamma^5}\tilde{\phi}_m(x)$,
being the matrix elements of the chiral rotation operator in the $a$-basis.
We can now simply express the Jacobian determinant in terms of the matrices $U$ and $\widetilde U$ as $\det J =\det \widetilde U U$.

We now reinterpret the Jacobian as a term in the action via the relation $\det J = e^{\tr\ln J}$.
Thus, our partition function becomes
\begin{equation*}
    Z
    = \int \mathcal{D}[\bar a', a']e^{-S'[\bar a', a'] - S_J},
\end{equation*}
where $S_J = \tr\ln J = \tr\ln\tilde U + \tr\ln U$.
Using the fact that $\dif{s}$ is infinitesimal we rewrite the above as
\begin{equation}
    S_J = -i \dif{s}\int \dif x (b \cdot x) \left[ I(x) + \tilde I(x)\right],
    \label{eq:Sj0}
\end{equation}
where
\begin{equation}
\begin{gathered}
    I(x) = \sum_n \phi^\dagger_n(x)\gamma^5\phi_n(x) \text{ and }
    \tilde{I}(x) = \sum_n \tilde{\phi}^\dagger_n(x)\gamma^5\tilde{\phi}_n(x).
\end{gathered}
\label{eq:Ifuncs}
\end{equation}

This is where one encounters an anomaly.
Let us first note that $I(x) + \tilde{I}(x)$ is exactly analogous to the anomaly function one encounters in computing the divergence of the axial current.
Now, the expressions in Eq.~\eqref{eq:Ifuncs} can readily be seen to be indeterminate.
Naively, from the completeness of the eigenstates $\phi$ and $\tilde \phi$ we have
\begin{equation*}
    I(x) = \tilde{I}(x) = \underbrace{\delta^{(4)}(0)}_{\infty} \times \underbrace{\tr \gamma^5}_{0}.
\end{equation*}
This ambiguity is due to the continuum representation of the path integral.
In order to resolve it we must introduce a proper regulator.
Following Refs.~[\onlinecite{Fujikawa1984,Zyuzin2012}], we evaluate $I$ and $\tilde{I}$ by heat kernel regularization as follows
\begin{multline}
    I(x) = \lim_{M\to\infty}
    \lim_{y \to x}
    \sum_n \phi^\dagger_n(y)\gamma^5e^{-\lambda_n^2/M^2} \phi_n(x)\\
     = \lim_{M\to\infty}
     \lim_{y \to x}
     \int_k \tr e^{i k \cdot y}\gamma^5e^{-\mathcal{D}^\dagger \mathcal{D}/M^2} e^{-ik\cdot x},
     \label{eq:heat}
\end{multline}
where we have used the completeness of the eigenfunctions $\phi_n(x)$ and $e^{-ik\cdot x}$.
The analogous expression holds for $\tilde{I}(x)$.
This has the benefit of regulating the expression in a gauge invariant manner.
The calculation of the anomaly functions $I$ and $\tilde I$ is presented in more detail in the appendix.
If we define the Hodge dual of the Maxwell tensor $\ast F^{\mu\nu} = (1/2)\epsilon^{\mu\nu\alpha\beta}F_{\alpha\beta}$ and the real and imaginary tensors $\Delta^{\mathrm{R}(\mathrm{I})}_{\mu\nu} = \Delta_{\mu\nu} \pm \Delta_{\mu\nu}^*$ we can express the result of the limit of $I + \tilde I$ in Eq.~\eqref{eq:heat} as
\begin{equation}
    -\frac{e^2}{8\pi^2} \ast\!{F}^{\mu\nu}F_{\mu\nu}
        + \frac{e}{4\pi^2} \mathrm{Re}\left[
            \ast{F}^{\mu\nu}\Delta^{\mathrm R}_{\mu\nu}
            + F^{\mu\nu}\Delta^{\mathrm I}_{\mu\nu}
    \right],
    \label{eq:anomalytensor}
\end{equation}
where we have assumed $b^\mu$ to be constant.

$\Delta_{\mu\nu}$ in Eq.~\eqref{eq:dirac} can be conveniently parametrized by the complex vector
\begin{equation}
g^i = \epsilon^{ijk} \Delta_{jk} + i \left(\Delta^{4i} - \Delta^{i4}\right),
\label{eq:g}
\end{equation}
where $i, j, k$ are spatial indices and the first and second terms are, respectively, purely real and imaginary due to Hermiticity.
This allows Eq.~\eqref{eq:anomalytensor} to be written in terms of vector quantities as
\begin{equation}
    I(x) + \tilde{I}(x)
    = -\frac{e^2}{2\pi^2} \mathbf{E} \cdot \mathbf{B}
    + \frac{e}{2\pi^2}\mathbf{E} \cdot \left(\mathbf{g}m^* + \mathbf{g}^* m\right).
    \label{eq:anomaly}
\end{equation}
In the above we have assumed $\mathbf{g}$ to have plane wave form.
The first term in Eq.~\eqref{eq:anomaly} is the conventional chiral anomaly, while the second term describes a new effect due to the added terms in the Lagrangian.

Note that the action in Eq.~\eqref{eq:Sj0} is linear in $\dif s$. We can thus perform a series of chiral gauge transformations to remove the axial vector $b^\mu$ from the electronic Lagrangian so that it becomes
\begin{equation}
    \mathcal{L} = \bar{\psi}'
    \left[
        i \slashed{D} - |m|e^{i(\alpha - 2 b\cdot x)\gamma^5}
        - \Delta_{\mu\nu}e^{-2ib \cdot x\gamma^5}
    \right]\psi'.
\end{equation}
This  corresponds to integrating the Jacobian in $s$ from $0$ to $1$. The Lagrangian arising from the Jacobian is
\begin{equation}
    \mathcal{L}_J =
    i b \cdot x\left[
        \frac{e^2}{2\pi^2} \mathbf{E} \cdot \mathbf{B}
        - \frac{e}{\pi^2}\mathbf{E} \cdot \mathrm{Re}\left[\mathbf{g}m^*\right]
    \right].
\label{eq:finallagrangian}
\end{equation}
Note that this result is unchanged if $m$ and $g$, instead of being constant, are taken to be plane waves with the same four-momentum, i.e.
\begin{equation}
    |m| \to |m| e^{-iQ \cdot x \gamma^5},\quad
    \Delta_{\mu\nu} \to \Delta_{\mu\nu} e^{-i Q \cdot x \gamma^5}.
\end{equation}
We may then, for example, take $Q = - 2b$.
In this case, the fermionic sector reduces to
\begin{equation}
    \mathcal{L} = \bar{\psi}'
    \left[
        i \slashed{D} - |m|e^{i\alpha\gamma^5}
        - \Delta_{\mu\nu}
    \right]\psi'.
\end{equation}
Since this has no dependence on $b^\mu$ we have isolated the axial vector dependent part of the current into the Jacobian term $\mathcal{L}_J$.
Alternatively, consider the case where $m$ and $\Delta$ arise from the decoupling of short range interactions in the appropriate channels.
In that case they will have Hubbard-Stratonovich Lagrangians
\begin{equation}
    \mathcal{L}_\text{HS} = \frac{1}{\lambda_m} |m(x)|^2 + \frac{1}{\lambda_\Delta} \Delta_{\mu\nu}(x) \Delta^{\mu\nu}(x).
\end{equation}
We can then absorb the chiral phase into the definitions of the parameters without affecting the Hubbard-Stratonovich term.

Now to illustrate the effects of Eq.~\eqref{eq:finallagrangian} let us consider the case of $\mathbf g = g_z \mathbf{\hat z}$.
After analytic continuation back to real time, the new term in the action reduces to
\begin{equation}
    S_{g} = \frac{e}{\pi^2} \int \dif x \  b \cdot x |m|g_z E_z \cos \alpha.
\end{equation}
This term has a form similar to the typical anomaly where in place of the magnetic field $\mathbf{B}$, we have $2\mathrm{Re}\, \left( {\mathbf{g} m^*} \right)$.
As such, we expect this term should also be capable of producing currents.
As there are no fermion operators in this term, we can obtain the induced currents by simply differentiating the action with respect to the vector potential.
Doing so we obtain a contribution
\begin{equation}
    \mathbf{j}_J = \frac{\delta S}{\delta \mathbf{A}} = \frac{e}{\pi^2}b_0 |m| g_z \cos\alpha \mathbf{\hat{z}}.
    \label{eq:current}
\end{equation}
We thus arrive at the prediction of a current without an external field.
This is best understood in a way akin to the chiral magnetic effect.
As has been discussed in a number of works, one can find zero or non-zero values for the chiral magnetic effect depending on the order of limits one uses in evaluating the result~\cite{Landsteiner2014,Chang2015,Alavirad2016}.
Taking the frequency to zero before momentum corresponds to the equilibrium case and as one would expect due to general arguments one finds no current in the absence of an electric field.
However, in the opposite limit, corresponding to a near equilibrium DC transport one finds that there is a current.
In the same way, one can interpret Eq.~\eqref{eq:current} as the response to a slow but non-zero frequency perturbation by the $\Delta_{\mu\nu}$ term in the action.


\begin{figure}[t!]
    \centering
    \includegraphics[width=0.75\linewidth]{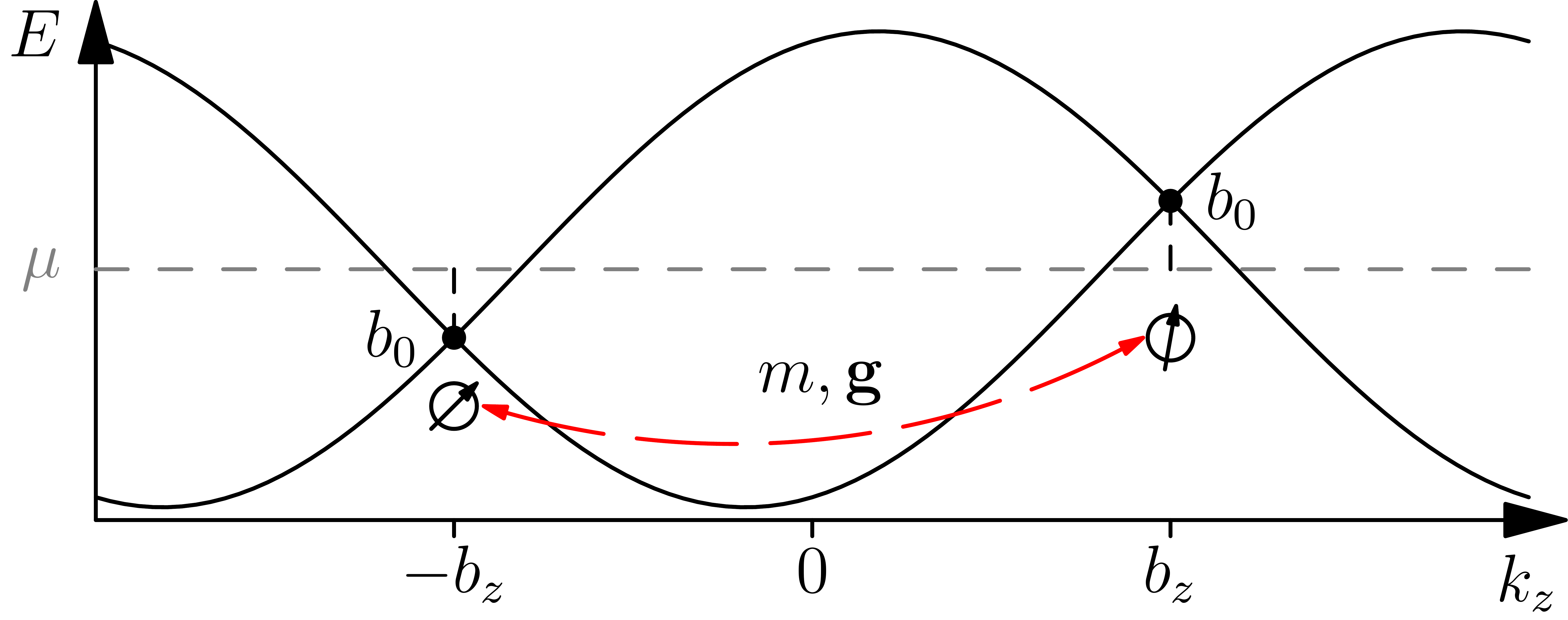}
    \caption{Two-band lattice model with a pair of Weyl nodes.
    The nodes are located at $(0, 0, \pm b_z)$ in the Brillouin zone and separated by an energy of 2$b_0$.\label{fig:weyl}}
\end{figure}

The ``enriched'' chiral anomaly as derived above is something only sharply defined for unbounded linearly dispersing particles~\cite{Burkov2015}.
In reality the Dirac theory of the previous section is only the low-energy description of some bounded dispersion in the Brillouin zone.
As such, we need to establish that the predicted effects can be observed within a lattice regulated model.

In order to verify the validity of the above conclusions independent of the subtleties of the low energy theory, we study the current response of a lattice model of Weyl-fermions.
Our purpose is to show that  the current response of the lattice system is in agreement with the prediction of the low energy theory, Eq.~\eqref{eq:current}.

\begin{figure}[t!]
    \centering
    \includegraphics[width=\linewidth]{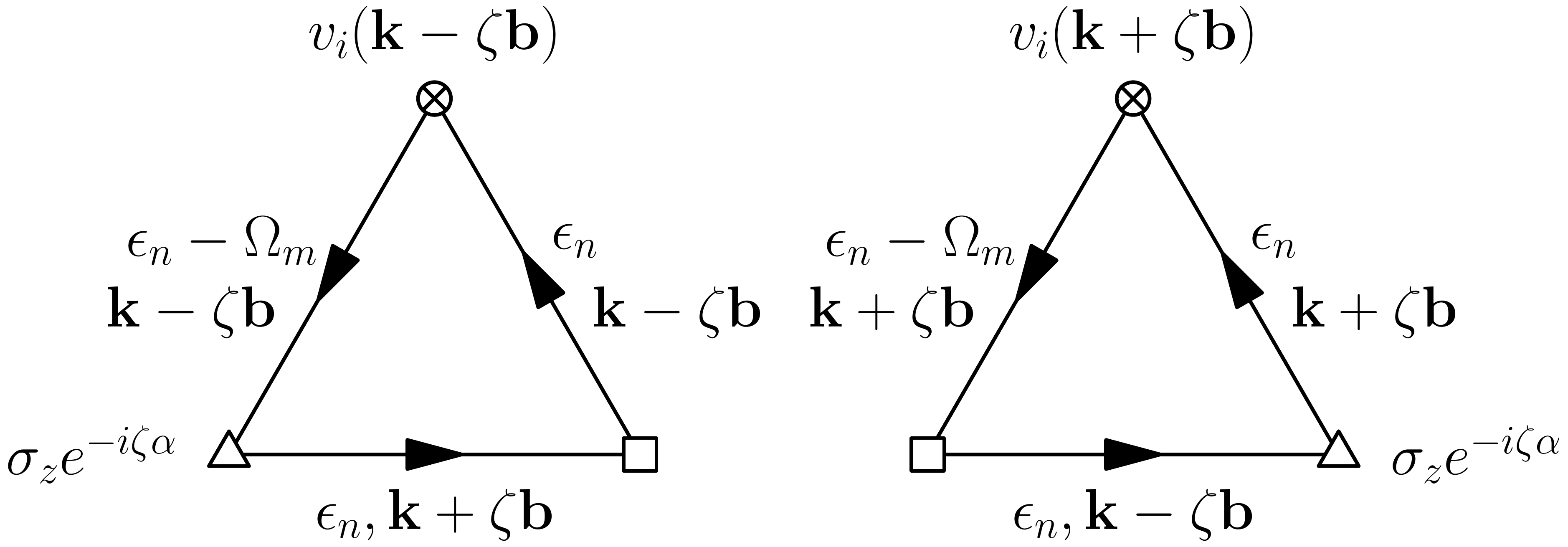}
    \caption{Diagrams contributing to the susceptibility $\chi_i$.
        The symbol $\otimes$ indicates the current vertex $v_i(\mathbf k)$, while the circle and square represent the vertices for $m$ and $g$ respectively.
        $\epsilon_n$ is the internal fermionic Matsubara frequency while $\Omega_m$ is the external bosonic Matsubara frequency to be analytically continued to obtain the retarded correlator.
        Internal loop momenta $\mathbf{k}$ and $\epsilon_n$ are summed over as well the index $\zeta=\pm1$.\label{fig:triangle}}
\end{figure}

In particular, we use the following inversion and time-reversal symmetry breaking two-band lattice model~\cite{Yang2011}
\begin{equation}
    H_0 = \sum_{\mathbf k} c^\dagger_\mathbf{k}
\left[
    \epsilon(\mathbf k) + \mathbf{d}(\mathbf k) \cdot \bm{\sigma}
\right]
c_{\mathbf k}
\label{eq:latticeH}
\end{equation}
with $\epsilon(\mathbf k) = t_1 \sin k_z$ and $\mathbf{d}^\dagger(\mathbf k) = \left(\sin k_x,\sin k_y,2 + \cos b_z - \sum_i\cos k_i \right)$,
which is host to a pair of Weyl fermions as depicted in Fig.~\ref{fig:weyl}. The momentum-space separation of the nodes is given by $2\mathbf{b} = 2b_z\mathbf{\hat z}$ and
the energy separation by $b_0 = 2t_1 \sin b_z$. To this bare Hamiltonian we add the perturbations
\begin{equation}
    H_m = m\sum_{\mathbf k} c^\dagger_{\mathbf k + \mathbf b}e^{-i\alpha}\sigma_z c_{\mathbf k - \mathbf b} + {\rm h.c.},
\end{equation}
which corresponds to the mass term in the low-energy theory of Eq.~\eqref{eq:dirac}.
The $\Delta_{\mu\nu}$ term can be modeled as
\begin{equation}
    H_g = \sum_{\mathbf k} c^\dagger_{\mathbf k + \mathbf b}\sigma_z \mathbf{g}(\tau)\cdot\bm{\sigma}c_{\mathbf k - \mathbf b} + {\rm h.c.}.
\end{equation}

We wish to establish the existence of a DC current in response to the combined terms $m$ and $\mathbf{g}$.
In particular, we calculate the retarded susceptibility of the current to $m$ and $\mathbf{g}$ in the uniform limit $\chi_{i}^R(\omega \to 0) = \lim_{\omega \to 0} \lim_{\mathbf q \to 0} \chi^R(\omega, \mathbf q)$.
$\chi^R$ is obtained as the analytic continuation from Matsubara frequency of the object
\begin{multline}
    \chi_{i}(i\Omega_m, \mathbf q) = \frac{\delta j_i(i \Omega_m, \mathbf q)}{\delta m \delta g(-i\Omega_m, -\mathbf q)}\\
    =  \left.\frac{\delta F[\mathbf{A}, g, m]}{\delta m \delta g(-i\Omega_m, \mathbf{q})\delta A_i(i\Omega_m, \mathbf{q})} \right|_{\substack{g=0\\ m=0\\ \mathbf{A}=0}},
        \label{eq:chi}
\end{multline}
where $F[\mathbf{A}, g, m]$ is the free energy in the presence of an external vector potential $\mathbf{A}$ and perturbations $m$, $g$.
Eq.~\eqref{eq:chi} corresponds to the diagrams in Fig.~\ref{fig:triangle} and describes the lowest order contribution of the $m$ and $g$ fields to the current in the spirit of linear response theory.

\begin{figure}[t!]
    \centering
    \includegraphics[width=\linewidth]{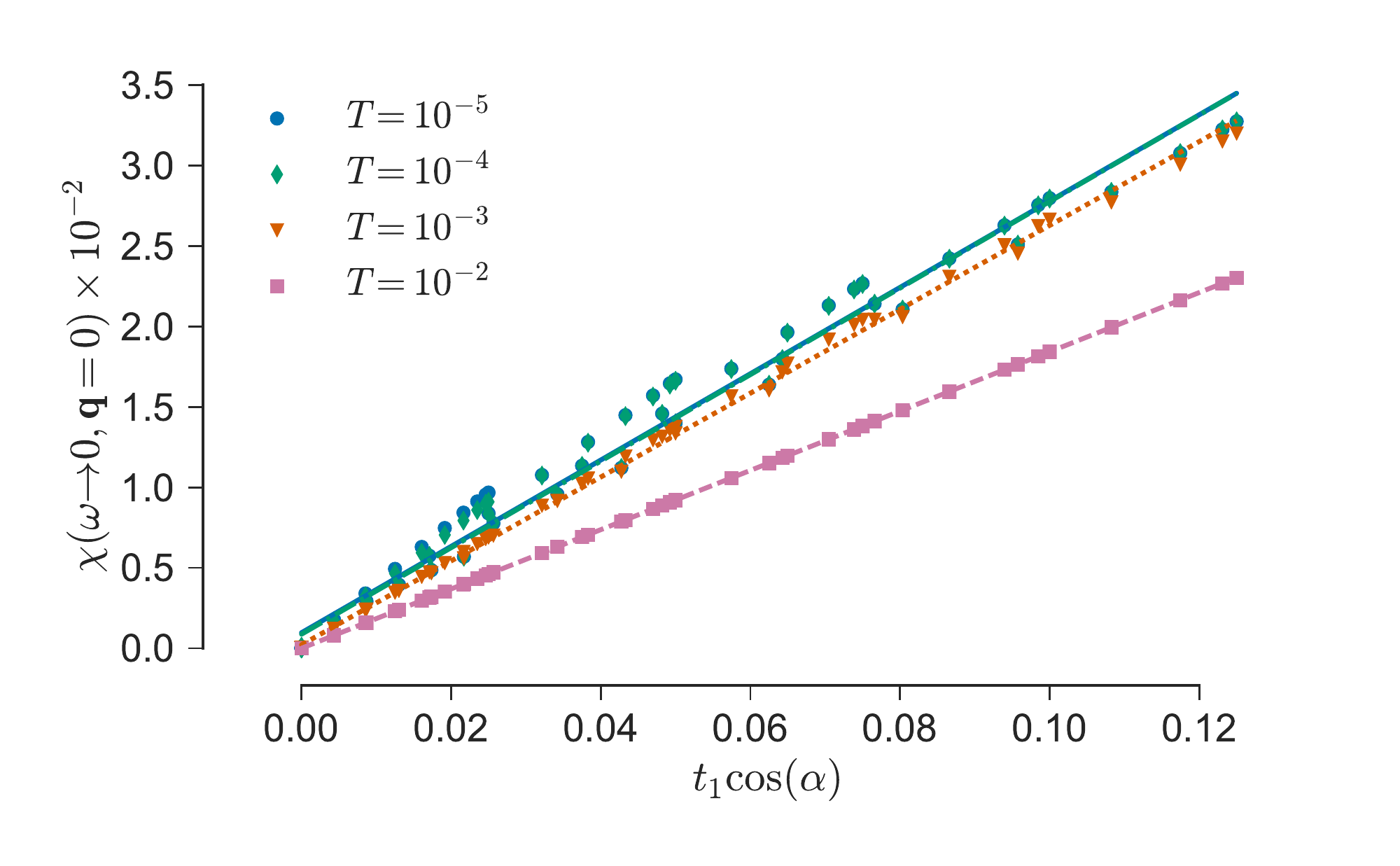}
 		\vspace*{-0.25in}
		\caption{The susceptibility $\chi$ in the uniform limit versus the parameter $t_1$ of Eq.~\eqref{eq:latticeH}.
        The induced current grows linearly with $t_1\cos\alpha = \frac{b_0}{\sin b_z}\cos\alpha$ as predicted by the low-energy theory, and vanishes in the absence of a nodal energy separation.\label{fig:chi}}
\end{figure}

As shown in Fig.~\ref{fig:chi}, the induced current grows linearly with the nodal energy separation and vanishes, as expected, when $t_1 = 0$.
We have also verified that in the opposite order of limits (with $\omega \to 0$ taken first), corresponding to the static equilibrium case, the current vanishes, as it must due to the Bloch's theorem for spontaneous currents~\cite{Bohm1949,Ohashi1996,Yamamoto2015}.

In terms of realizing the effect discussed above, one may note that Eq.~\eqref{eq:dirac} can be interpreted as a mean-field model subject to interactions in the appropriate channels.
Without much generalization we can write
\begin{multline}
    \mathcal{L} = \mathcal{L}_0 - \lambda_m \bar{\psi}e^{i\alpha\gamma^5}\psi \bar{\psi}e^{-i\alpha\gamma^5}\psi
    - \lambda_\Delta\bar{\psi}\sigma^{\mu \nu}\psi\bar{\psi}\sigma_{\mu \nu}\psi.
    \label{eq:interacting}
\end{multline}
A Hubbard-Stratonovich decoupling, then leads to a spacetime-dependent analog of Eq.~\eqref{eq:dirac}.
In a condensed matter context we can write Eq.~\eqref{eq:interacting} in terms of spin, $\bm{\sigma}$, and valley, $\tau_i$, degrees of freedom as
\begin{multline}
   \mathcal{L} = c^\dagger(\partial_\tau + \mathbf{k} \cdot \bm{\sigma}\tau_z - \mathbf{b} \cdot \bm{\sigma} - b_0 \tau_z)c\\
    - \lambda_m e^{i\alpha} \left(\bar{c}\tau_+ c\right) \left(\bar{c} \tau_- c\right)
    - \lambda_\Delta \left(\bar{c} \mathbf{g} \cdot \bm{\sigma}\tau_+ c\right)\left( \bar{c} \mathbf{g}^* \cdot \bm{\sigma} \tau_- c\right).
    \label{eq:intlattice}
\end{multline}
We therefore interpret $m$ as a charge-density wave, while $\mathbf{g}$ describes a spin-density wave, since the valley degree of freedom denotes a separation in momentum space.
This suggests that our model may potentially be realized in a system of interacting Weyl electrons, where interactions give rise to such density-wave orders. Alternatively,
such perturbations may be externally induced, e.g.\ in Floquet-driven Weyl materials.

\begin{acknowledgments}
This work was supported by U.S. Department of Energy {BES-DESC0001911} and Simons Foundation.
\end{acknowledgments}

\bibliography{references}

\appendix*
\section{Supplemental information: Heat kernel regularization of the anomaly functions}
\label{sec:regulate}

We wish to evaluate the heat kernel regulated expressions
\begin{equation}
    I(x) = \lim_{M\to\infty}
     \lim_{y \to x}
     \int_k \tr e^{i k \cdot y}\gamma^5e^{-\mathcal{D}^\dagger \mathcal{D}/M^2} e^{-ik\cdot x}.
     \label{eq:ah1}
\end{equation}
and
\begin{equation}
    \tilde{I}(x) = \lim_{M\to\infty}
     \lim_{y \to x}
     \int_k \tr e^{i k \cdot y}\gamma^5e^{-\mathcal{D}\mathcal{D}^\dagger/M^2} e^{-ik\cdot x}.
     \label{eq:ah2}
\end{equation}

To do so, let us first define the operator
\begin{equation}
    O = i |m| e^{i \alpha \gamma^5} + i \Delta_{\mu\nu} \sigma^{\mu\nu}.
\end{equation}
We then expand
\begin{multline}
    \mathcal{D}^\dagger \mathcal{D} = -D^2 + \frac{e}{2} F_{\mu\nu}\sigma^{\mu\nu}
    + i [\slashed{D}, \slashed{b}]\gamma^5\\
    + \slashed{D}O + O^\dagger\slashed{D}
    + i\slashed{b}\gamma^5 O + iO^\dagger \slashed{b} \gamma^5
    + O^\dagger O - b^2
\end{multline}
and similarly
\begin{multline}
    \mathcal{D}\mathcal{D}^\dagger = -D^2 + \frac{e}{2} F_{\mu\nu}\sigma^{\mu\nu}
    + i [\slashed{D}, \slashed{b}]\gamma^5\\
    + O\slashed{D} + \slashed{D}O^\dagger
    + O i\slashed{b}\gamma^5 + i \slashed{b} \gamma^5O^\dagger
    + OO^\dagger - b^2.
\end{multline}
where $X^2$ denotes $\sum_i X_i^2$.
Here, we have used the identity
\[
    [D_\mu, D_\nu] = ie F_{\mu\nu},
\]
where $F_{\mu\nu}$ is the electromagnetic field tensor.
The presence of the plane waves in Eqs.~(\ref{eq:ah1}) and~(\ref{eq:ah2}), only leads to the replacement $D_{\mu} \to D_{\mu} - i k_\mu$.
We thus can rewrite Eq.~\eqref{eq:ah1} as
\begin{multline}
    I(x) =
     \lim_{M\to\infty}
     M^4\int_p e^{-p^2}\tr \gamma^5\exp\left[ -\frac{\mathcal{D}^\dagger \mathcal{D}}{M^2}\right.\\
       \left.  + M^{-1}\left( - i p^\mu D_\mu +  [\slashed{p}, \slashed{b}]\gamma^5 - i\slashed{p} O - i O^\dagger \slashed{p}
         \right)     \right],
\end{multline}
where we have also made the change of variables $p^\mu = M k^\mu$.
Writing the exponential term as
\[
    \exp\left(-\frac{\mathcal{D}^\dagger \mathcal{D}}{M^2} + \frac{K}{M}\right),
\]
we can expand in powers of $M$
\begin{multline}
    I(x) = \lim_{M\to\infty}
     M^4\int_p e^{-p^2}\\
     \times
     \tr \gamma^5\left[1
         + \frac{K}{M}
         + \frac{K^2 - 2\mathcal{D}^\dagger \mathcal{D}}{2M^2}\right.\\
         \left.
         + \frac{K^3/6 - \{K, \mathcal{D}^\dagger \mathcal{D}\}}{M^3}
         + \frac{K^4/4! + \mathcal{D}^\dagger \mathcal{D}^2/2}{M^4}
     \right].
\end{multline}

As in the conventional case, the only terms which survive are at order $\frac{1}{M^4}$.
All lower-order or $K$-dependent terms either vanish due to the matrix traces or cancel between $I$ and $\tilde I$.
We are then left with
\begin{equation}
    I(x) + \tilde{I}(x) = \frac{1}{2} \tr\left[{(\mathcal{D}^\dagger \mathcal{D})}^2 + {(\mathcal{D}\mathcal{D}^\dagger)}^2\right]\int \frac{\dif^4\!{p}}{{(2\pi)}^4} e^{-p^2}.
\end{equation}
We can easily perform the Gaussian integral to obtain a factor of $\pi^2$.
Computing the matrix traces we arrive at Eq.~\eqref{eq:anomaly}.

\end{document}